\documentclass[
reprint,
amsmath, amssymb,
pra,
aps,
superscriptaddress, twocolumn,nofootinbib,groupedaddress,
]{revtex4-2}

\usepackage{graphicx}
\usepackage{subfigure}
\usepackage{hyperref}
\usepackage{bm}
\usepackage{float}
\usepackage{color}
\usepackage{amsmath}
\usepackage{braket} 
\usepackage{ulem}

\hypersetup{colorlinks=true, citecolor=blue, urlcolor=blue, linkcolor=blue}
\begin{document}
\title{Non-reciprocal Synchronization in Thermal Rydberg Ensembles}

\author{Yunlong Xue$^{1}$}
\author{Zhengyang Bai$^{2}$}
\email{zhybai@nju.edu.cn}

\affiliation{
$^1$State Key Laboratory of Precision Spectroscopy, East China Normal University, Shanghai 200062, China\\
 $^2$National Laboratory of Solid State Microstructures and School of Physics,
Collaborative Innovation Center of Advanced Microstructures, Nanjing University, Nanjing 210093, China}
\date{\today}


\begin{abstract}
Optical non-reciprocity is a fundamental phenomenon in photonics. It is crucial for developing devices that rely on directional signal control, such as optical isolators and circulators. However, most research in this field has focused on systems in equilibrium or steady states. 
In this work, we demonstrate a room-temperature Rydberg atomic platform where the unidirectional propagation of light acts as a switch to mediate time-crystalline-like collective oscillations through atomic synchronization.
We find that thermal-motion-induced coupling asymmetry, enabled by counter-propagating probe and control fields, generates persistent oscillations; conversely, co-propagation quenches this effect.
We identify, through both numerical and analytical approaches, the criteria for realizing optical non-reciprocity within a synchronization regime. These results provide key insights for chiral quantum optics and promote the on-chip integration of non-reciprocal devices in nonequilibrium many-body systems.
\end{abstract}
\maketitle

\textit{Introduction}.---Non-reciprocal optical components are cornerstones of modern photonic chips and quantum information networks~\cite{dutt2024Nonlinear,luo2023Recent,kim2021Onchip,luo2019Nonlinear,sounas2017Nonreciprocal,shen2016Experimental,khanikaev2015Nonlinear,lodahl2017Chiral}.
Traditional implementations have been mainly dependent on magneto-optical effects~\cite{huang2020Tuning,feng2020Topological}, whereas recent advances have expanded to emerging areas such as nonlinear and quantum optics~\cite{wang2025SelfinducedOptical,liang2020CollisionInduced,zhang2018Thermalmotioninduced,shen2016Experimental,huang2018Nonreciprocal}. 
A particularly promising approach utilizes the light propagation direction itself as a switching degree to control macroscopic responses \cite{zola2019Dynamic,dorrah2021Metasurface,jin2018IncidentDirection}.
In thermal atomic vapors, it has been demonstrated that a unidirectional control field, through interaction with atomic thermal motion, induces susceptibility–momentum locking via the Doppler effect, yielding pronounced steady-state non-reciprocity~\cite{zhang2018Thermalmotioninduced, liang2020CollisionInduced}. This strategy effectively harnesses thermal motion as a resource. It exploits direction-dependent light–matter interactions to break optical symmetry, thereby enabling precisely controllable responses. 

A natural question arises as to whether such direction-dependent control mechanisms can also manipulate the dynamics of optical fields in nonequilibrium states and facilitate chiral switching in quantum many-body systems~\cite{xie2025Chiral, wadenpfuhl2023EmergenceSynchronization,wu2024Dissipative,ding2024Ergodicity}. 
Exploring these chiral nonequilibrium effects demands an optical medium capable of sustaining strong and tunable light–matter interactions. Rydberg atomic gases inherently satisfy this requirement. When excited to high-lying Rydberg states, atoms exhibit strong dipole–dipole interactions, effectively converting the ensemble into a highly correlated quantum many-body medium \cite{bernien2017Probing,adams2020Rydberg,bendkowsky2009Observation,saffman2010Quantum}.
These interactions give rise to rich many-body phenomena, such as antiferromagnetic phases~\cite{bernien2017Probing, ebadi2021Quantuma, guardado-sanchez2018Probing}, quantum scars~\cite{turner2018Weak, bluvstein2021Controlling}, and many-body synchronization~\cite{wadenpfuhl2023EmergenceSynchronization,wu2024Dissipative,ding2024Ergodicity}.

\begin{figure}[h] 
\includegraphics[width=0.45\textwidth]{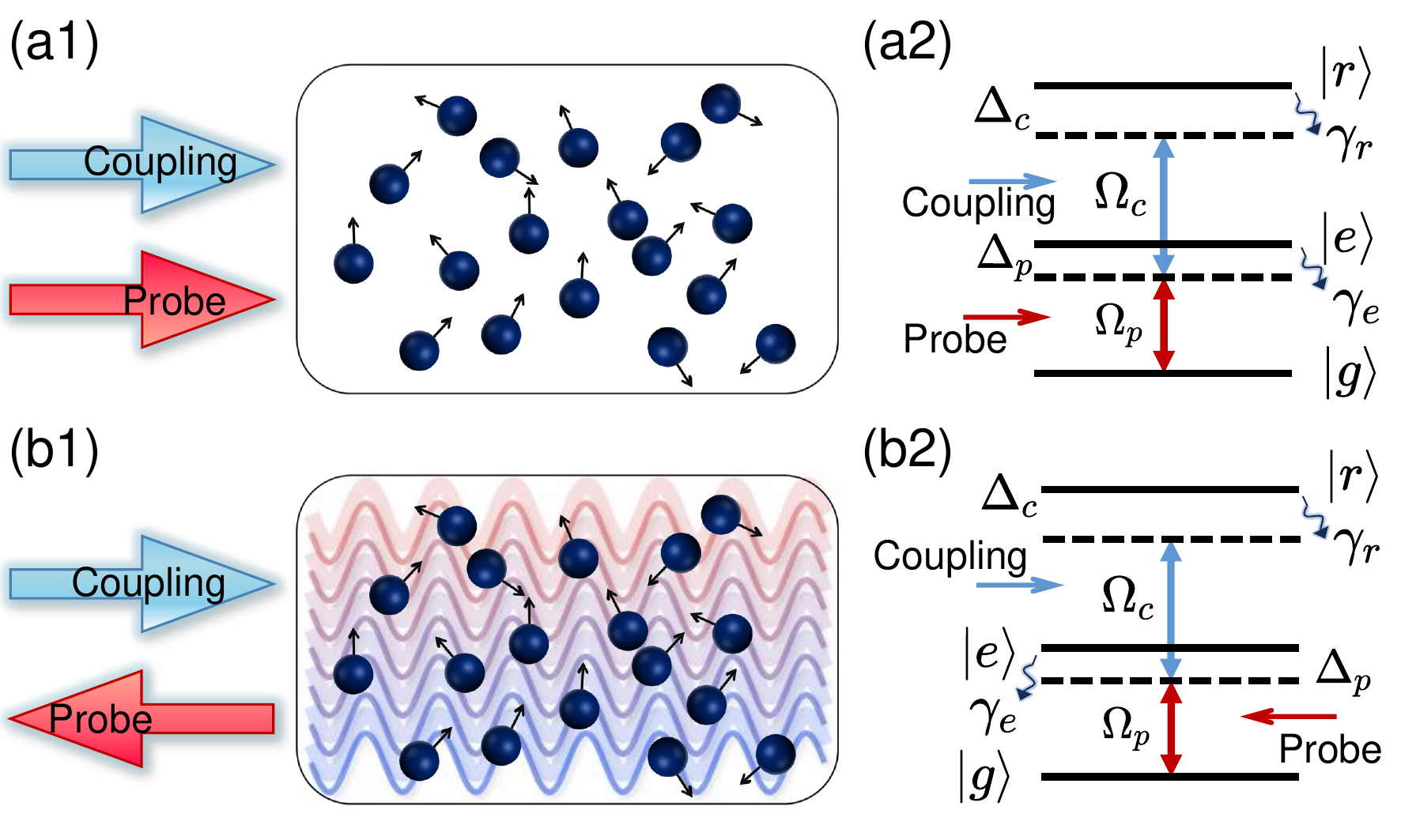} 
\caption{\textbf{Schematic diagram of the optical non-reciprocal synchronization model.} (a1) and (b1) Schematics of the system response under co- and counter-propagating probe and coupling fields, respectively. (a2) and (b2) The corresponding energy-level diagram of the three-level Rydberg atom for each configuration. } \label{fig:1} 
\end{figure}

In this work, we propose a theoretical scheme for achieving optical non-reciprocity in nonequilibrium states using thermal Rydberg ensembles.
Our results show that in the counter-propagating configuration, atoms with different velocities synchronize via Rydberg interactions, leading to oscillatory dynamical phases characterized by frequency and phase locking in a subset of atoms.
Conversely, both synchronization and the ensuing phases are suppressed under co-propagation. This establishes a direct link between optical non-reciprocity and nonequilibrium phases, thereby enabling the optical propagation direction to act as a switch for quantum dynamical states. 
Beyond its fundamental applications, 
the ability to operate at room temperature paves the way for designing and realizing practical nonequilibrium quantum optics devices.

\textit{Model}.---We consider an ensemble of three-level Rydberg atoms driven by external probe and coupling fields, as illustrated in Fig.~\ref{fig:1}(a2)[(b2)]. Atoms comprise a ground state $|g\rangle$, an intermediate excited state $|e\rangle$, and a highly excited Rydberg state $|r\rangle$. After applying a transformation to the rotating frame with respect to the free Hamiltonian and making the rotating-wave approximation, the atom–light interaction Hamiltonian takes the following form ($\hbar=1$),
\begin{equation}
\begin{aligned}
\hat{H}_{\rm{ap}} = &-\Delta_p \hat\sigma_{ee} - (\Delta_p + \Delta_c) \hat\sigma_{rr} \\
         & + \frac{1}{2} \left( \Omega_p \hat\sigma_{eg}+\Omega_c \hat\sigma_{re}+\rm{H.c.}  \right),
\end{aligned}
\end{equation}
where $\hat{\sigma}_{\alpha\beta}$
are atomic transition operators ($\alpha, \beta \in \{g, e, r\}$). Here, $\Omega_p$ and $\Omega_c$ represent the Rabi frequencies of the probe and coupling lasers, respectively, while $\Delta_p$ and $\Delta_c$ denote their corresponding detunings.  

\begin{equation}
    \hat H_{\rm{int}} =-\mathcal{N}\int d^{3}\mathbf{r}^{\prime}\hat{\sigma}_{rr}\left(\mathbf{r}^{\prime},t\right)V_{rr}(\mathbf{r}-\mathbf{r}^{\prime})\hat{\sigma}_{rr}\left(\mathbf{r},t\right).
\end{equation}
Typically, the interaction between Rydberg atoms at positions $\mathbf{r}$ and $\mathbf{r}^{\prime}$ is described by the van der Waals potential $V_{rr}(\mathbf{r}-\mathbf{r}^{\prime}) = C_6 / |\mathbf{r}-\mathbf{r}^{\prime}|^6$, where $C_6$ is the dispersion coefficient, and $\mathcal{N}$ denotes the atomic density.

The system is described microscopically using a quantum master equation for the density operator $\hat{\rho}$,
\begin{equation}\label{eq:master}
    \frac{d\hat{\rho}}{dt} = -i[\hat{H},\hat{\rho}] + \mathcal{N}\sum_{\alpha=r, e} \gamma_{\alpha}\int d^{3}{\mathbf r} \left( \hat{L}_\alpha \hat{\rho} \hat{L}_\alpha^\dagger - \frac{1}{2} \left\{ \hat{L}_\alpha^\dagger \hat{L}_\alpha, \hat{\rho} \right\} \right),
\end{equation}
where the Hamiltonian $\hat{H}$ responsible for the coherent dynamics reads
$\hat{H}=\mathcal{N}\int d^3r [\hat{H}_{\rm{ap}}(\mathbf{r}, t)+\hat H_{\rm{int}} (\mathbf{r}, t)]$. 
The second term on the right-hand side of Eq.~(\ref{eq:master}) describes the dissipation process. In the Lindblad term,  the operator $\hat{L}_r=\hat{\sigma}_{er}({\mathbf r}, t)$ and $\hat{L}_e=\hat{\sigma}_{ge}({\mathbf r}, t)$ corresponds to the transition $|r\rangle\to|e\rangle$ and $|e\rangle\to|g\rangle$ with the decay rate $\gamma_r$ and $\gamma_e$. The anticommutator is defined as $\{\hat{A}, \hat{B}\} = \hat{A}\hat{B} + \hat{B}\hat{A}$.

\begin{figure}[htp]
 \includegraphics[width=0.45\textwidth]{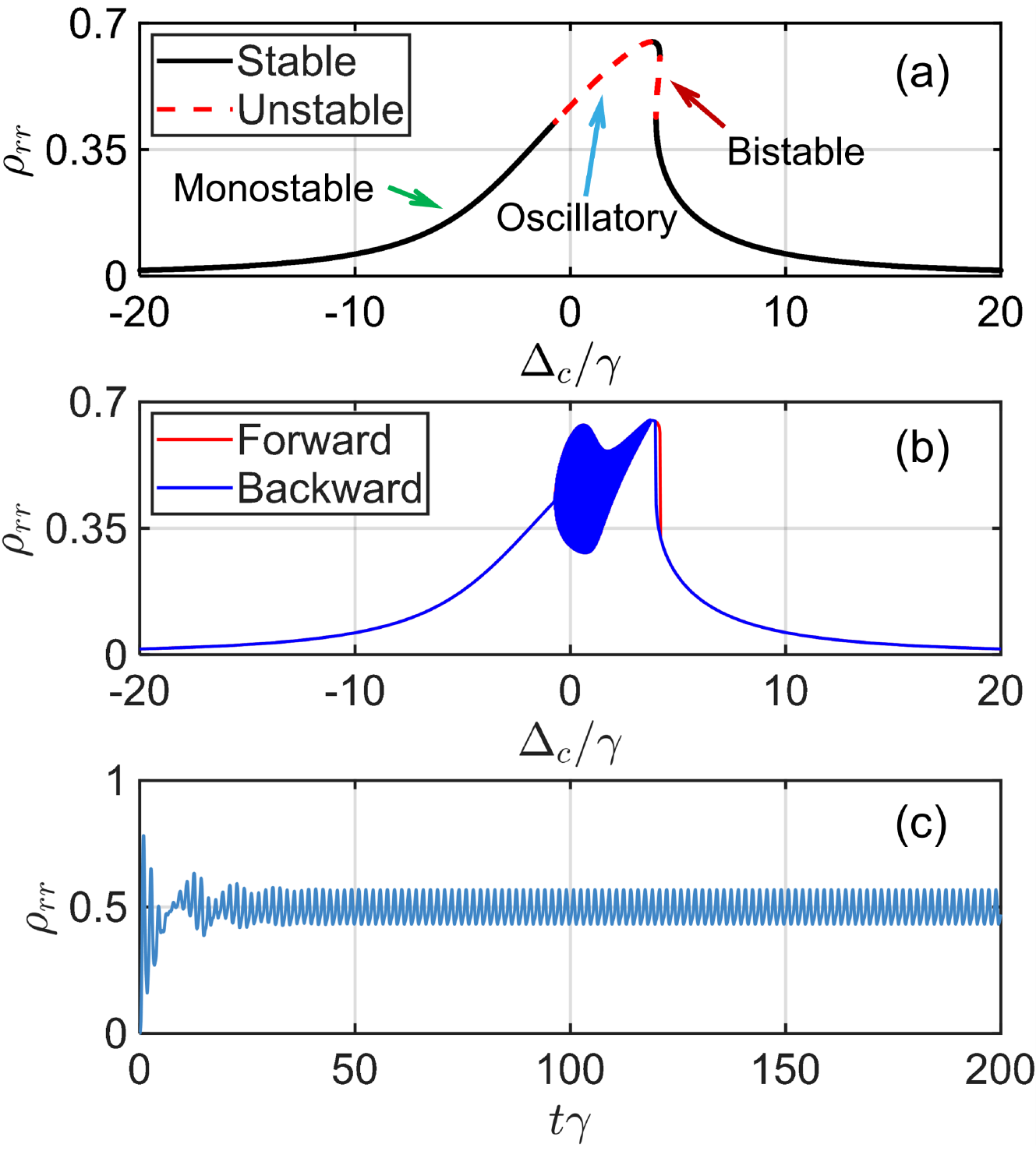}
    \caption{{\bf The stability analysis and its dynamical simulation.} 
(a)  As detuning $\Delta_c$ is varied, the system is partitioned into three distinct regimes based on the number and stability of fixed points: monostable, bistable, and oscillatory.
(b) We adiabatically sweep the $\Delta_c$ in both directions to measure the values of $\rho_{rr}$.
(c) Time dynamics of $\rho_{rr}$ in the oscillatory regime at $\Delta_c=0$.
Simulation parameters are set as follows: $b=2$, $\gamma_{e} = \gamma = 1$, $\gamma_{r} = 10^{-3}\gamma$, $\Delta_p = 0$, $\overline{V}_{rr} = -9\gamma$, $\Omega_{c} = 4.4\gamma$, and $\Omega_{p} = 6\gamma$.}
\label{fig:2}
\end{figure}

Due to the dissipative nature of the Rydberg ensemble, mean-field (MF) theory can be employed to characterize the dynamical behavior of the system. 
Within this approximation, the many-body density matrix is decoupled into a product of single-particle matrices, which effectively neglects spatial correlations between different sites. This approximation adequately captures the dynamical behavior of the system as it approaches the thermodynamic limit, and provides a good description in high dimensions, where fluctuations from neighboring sites tend to average out~\cite{lee2011AntiferromagneticPhase, carr2013Nonequilibrium}.

\begin{figure*}[htp]
 \includegraphics[width=1\textwidth]{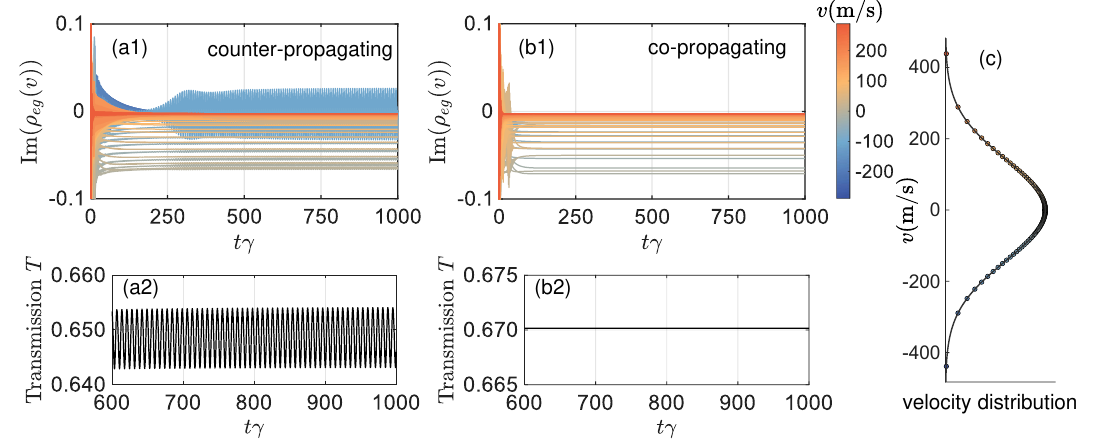}
 \caption{{\bf Non-reciprocal synchronization dynamics.} (a1)[(b1)] Dynamics of $\mathrm{Im} [\rho_{eg}(v)]$ for atomic velocities between $-400$ m/s and $400$ m/s under the counter-propagating (co-propagating)  condition. (a2)[(b2)] Dynamics of transmission $T$ under co-propagating (counter-propagating) conditions after Doppler averaging. (c) The velocities of the 150 atomic groups used in the simulation are initialized by sampling from the Maxwell-Boltzmann distribution at a temperature of $T_c = 321~\mathrm{K}$. Simulation parameters are set as follows: $\Omega_p = 6\,\gamma$, $\Omega_c = 4\,\gamma$, $\Delta_p = 0$, $\Delta_c = -11\,\gamma$, and $\overline{V}_{rr} = 800\,\gamma$.}
\label{fig:3}
\end{figure*} 

Starting from the Lindblad master equation (\ref{eq:master}), we derive the optical Bloch equations for the density matrix elements
$\rho_{\alpha\beta}({\bf r},t) \equiv \langle \hat{\sigma}_{\alpha\beta}({\bf r},t) \rangle$, for example, the population dynamics are
\begin{subequations}\label{eq;population}
    \begin{eqnarray}
        \frac{d\rho_{gg}}{dt} 
        &=&\gamma_e \rho_{ee} - \tfrac{i}{2}\left(-\Omega_{p}\rho_{ge} + \Omega_{p}\rho_{eg}\right), \\
        \frac{d\rho_{ee}}{dt}
        &=& -\gamma_e \rho_{ee}+\gamma_r\rho_{rr}\nonumber\\&& - \tfrac{i}{2}\left[\Omega_{p}(\rho_{ge}-\rho_{eg})+ \Omega_{c}(\rho_{re} - \rho_{er}) \right] ,\\
        \frac{d\rho_{rr}}{dt} 
        &=&-\gamma_r \rho_{rr} - \tfrac{i}{2}\left(\Omega_{c}\rho_{er} - \Omega_{c}\rho_{re}\right),\\
        \frac{d\rho_{ge}}{dt}
        &=& d_{ge} \rho_{ge} - \tfrac{i}{2}\left(-\Omega_{p}\rho_{gg} - \Omega_{c}\rho_{gr} + \Omega_{p}\rho_{ee}\right) ,\\
        \frac{d\rho_{gr}}{dt} 
        &=& d_{gr} \rho_{gr} - \tfrac{i}{2}\left(-\Omega_{c}\rho_{ge} + \Omega_{p}\rho_{er}\right)\nonumber \\
        &-&i\overline{V}_{rr}\rho_{rr}^b\rho_{gr},\label{eq;population-e} \\
        \frac{d\rho_{er}}{dt}
        &=&d_{er} \rho_{er} - \tfrac{i}{2}\left(\Omega_{p}\rho_{gr} - \Omega_{c}\rho_{ee} + \Omega_{c}\rho_{rr}\right) \nonumber\\
        &-&i\overline{V}_{rr}\rho_{rr}^b\rho_{er},\label{eq;population-f}
    \end{eqnarray}
\end{subequations}
where $d_{\alpha\beta}=i(\Delta_{\alpha}-\Delta_{\beta})-\gamma_{\alpha\beta}$, with $\gamma_{\alpha\beta}\equiv(\gamma_\alpha+\gamma_\beta)/2$, and $\overline{V}_{rr}$ represents the effective MF interaction strength. 
In Eqs.~(\ref{eq;population-e}) and (\ref{eq;population-f}), we introduce a MF shift of the Rydberg level that depends on the Rydberg density via a power law, $\overline{V}_{rr} \rho_{rr}^b$, parameterized by an exponent $b$.
This dependence arises because Rydberg atoms interact via a power-law potential. Within the mean-field theory, this leads to an effective shift of the Rydberg level that scales with the density as $\rho_{rr}^b$~\cite{wadenpfuhl2023EmergenceSynchronization}.

Importantly, the steady-state solutions of Eqs.~\eqref{eq;population} are not always unique due to the nonlinearity induced by Rydberg interactions. 
We characterize the phases using the Rydberg population $\rho_{rr}$ as an order parameter.
Through linear stability analysis around the fixed points, three characteristic dynamical regimes are identified: the monostable phase, which has a single stable steady state; the bistable phase, characterized by two coexisting stable solutions; and the oscillatory phase, in which no stable stationary fixed point exists [Fig.~\ref{fig:2}(a)].

To probe the dynamical phase of the system, we sweep the detuning $\Delta_c$ adiabatically from large positive to large negative values, and vice versa [Fig.~\ref{fig:2}(b)]. Near resonance, the system exhibits rapid oscillations due to the absence of stable fixed points. This interplay between driving and dissipation drives the system into a time-crystalline phase~\cite{wu2024Dissipative}. Within the bistable regime, the forward and backward sweeps follow distinct trajectories, resulting in a hysteresis loop~\cite{carr2013Nonequilibrium}.

Physically, the transition from the monostable to the bistable phase stems from the interplay between coherent driving and nonlinear Rydberg interactions, whereas the oscillatory regime arises  when this competition destabilizes all stationary solutions. Remarkably, the time-crystalline phase of interest here naturally arises within this non-stationary regime, where the system maintains persistent oscillations in the absence of any stable steady state. As shown in Figs.~\ref{fig:2}(c), this is accompanied by a breaking of time-translation symmetry, dynamically manifested as sustained and undamped oscillations. 

\textit{Non-reciprocal Synchronization}.---Inspired by the oscillatory dynamics observed in Rydberg systems, we will extend our research to room-temperature (or thermal) Rydberg gases. This will allow us to study synchronization among atoms belonging to different velocity classes and pursue optical non-reciprocity in nonequilibrium states.

To simulate the dynamics in thermal atomic systems, Doppler shifts arising from atomic motion are accounted for: atoms with velocity $v$ experience modified detunings $\Delta_\alpha \rightarrow \Delta_\alpha + \mathbf{k}_\alpha \cdot \vec{v}$ for $\alpha = p, c$, where $\mathbf{k}_p$ ($\mathbf{k}_c$) is the wave vector of the probe (coupling) field. Assuming the system is in thermal equilibrium, the atomic velocities follow the Maxwell–Boltzmann distribution
$f(v) = 1/(\sqrt{\pi} v_{T_c}) {\rm exp}{-(v/v_{T_c})^2}$, where $ \quad v_{T_c} = \sqrt{2k_B T_c / M}$ is the most probable speed at temperature $T_c$, and $M$ denotes the atomic mass. Then, the macroscopic response can be obtained by thermal averaging over velocity-dependent density matrix elements
$\tilde{\rho}_{\alpha\beta} = \int dv\, f(v) \rho_{\alpha\beta}(v)$. 

Such Doppler broadening is a common feature in room-temperature atomic gas experiments and typically averages out nontrivial dynamical features in conventional quantum many-body systems, making them effectively undetectable in macroscopic measurements. However, in our system, atoms belonging to different velocity classes are nonlinearly coupled through MF interactions of the form
$\int \mathrm{d}v f(v)\overline{V}_{rr}\rho_{rr}^b(v)\rho_{\alpha\beta}(v)$
($\alpha \neq \beta$; $\alpha, \beta = g, e, r$).
As a result, the evolution of each velocity class is not independent of the others. This is the essential element of the atomic synchronization mechanism~\cite{bai2020SelfInduced, wadenpfuhl2023EmergenceSynchronization}.

As illustrated in Fig.~\ref{fig:3}(a1), we analyze the dynamical behavior of the system across different velocity classes, with special emphasis on Doppler broadening effects. It is found that dynamics of atoms with different velocity classes are largely different during the initial transient period. At a later stage ($t\gamma> 200$),  
we observe the emergence of synchronization as a significant fraction of atoms from different velocity classes are attracted toward a limit cycle, where they begin to oscillate with a common frequency and fixed phase relationship. As more velocity classes become phase-locked, the MF strength increases. This, in turn, drives additional velocity classes to align their oscillations, ultimately leading to a partially or fully synchronized state. In the simulation, we generate $150$ groups of atoms with distinct velocities, whose statistical weights are shown in Fig.~\ref{fig:3}(c) and follow a Maxwell--Boltzmann distribution. The interaction density term is updated iteratively at each time step.

After Doppler averaging, the emergence of atomic synchronization leads to persistent, time-periodic oscillations in the optical transmission as the system evolves toward a nonequilibrium steady state [Fig.~\ref{fig:3}(a2)]. This behavior represents a hallmark signature of the spontaneous breaking of temporal translational symmetry. The optical transmission is given by $T={\rm exp}[-k_p{\rm Im}(\chi_p) L]$, where the susceptibility  $\chi_p=2{\cal N} |\mu_{eg}|^2\int dvf(v)\rho_{ge}(v)/[\varepsilon_0\Omega_p]$ integrates over the velocity distribution. Here, $\mu_{eg}$ denotes the dipole moment between the ground and excited states, and $L$ is the length of the optical medium.
Therefore, the emergence of macroscopic oscillations in the optical response of a hot Rydberg vapor is a manifestation of a Kuramoto-like synchronization transition~\cite{kuramoto1975international, acebron2005Kuramoto}, which is driven by sufficiently strong Rydberg interactions.  
Such synchronized oscillations correspond to the oscillatory dynamics recently observed in experiments~\cite{wadenpfuhl2023EmergenceSynchronization,wu2024Dissipative,ding2024Ergodicity}.

The coupled Eq.~\eqref{eq;population} exhibits motion-induced non-reciprocity due to Doppler shifts in thermal Rydberg gases~\cite{lodahl2017Chiral,zhang2018Thermalmotioninduced}. In simulations of the synchronized state, the coupling and probe fields are counter-propagating to suppress the first-order Doppler broadening in the two-photon process.
However, under co-propagating illumination with a reversed incidence direction, atoms across the velocity distribution exhibit a conventional dynamical response [Fig.~\ref{fig:3}(b1) and (b2)]. The optical transmission undergoes rapid transient oscillations prior to settling into a steady state. This stands in sharp contrast to the synchronized behavior seen in Fig.~\ref{fig:3}(a1) and (a2). 
Notably, chiral coupling is induced by atomic thermal motion. This coupling causes the system to exhibit markedly different dynamical responses under distinct propagation conditions, thereby realizing a non-reciprocal synchronized state.

Similar to the Kuramoto model~\cite{wadenpfuhl2023EmergenceSynchronization,acebron2005Kuramoto}, a time-crystal phase emerges via synchronization among atoms of different velocity classes. 
The synchronization is governed by each group's intrinsic frequency, the coupling strength, and the power and detuning of the driving fields. Crucially, the propagation direction of the laser field modifies the effective detuning distribution across velocity groups. This modification disrupts the synchronization parameters, thereby causing the different dynamics observed under different propagation directions.

\begin{figure}[htp]
 \includegraphics[width=0.50\textwidth]{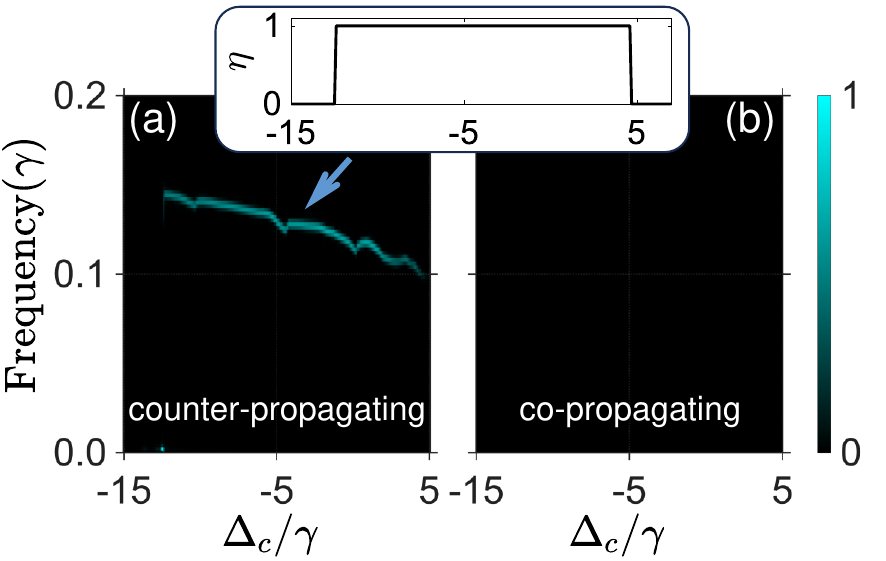}
  \caption{{\bf Non-reciprocal synchronized response versus the detuning of the coupling field. } (a)[(b)] The oscillation frequencies of  transmission under counter-propagating (co-propagating) condition. The color map represents the normalized oscillation amplitude. The simulation parameters are consistent with those in Fig.~\ref{fig:3}. The inset shows the contrast ratio $\eta$ versus $\Delta_c$. }
\label{fig:4}
\end{figure} 

We further test the robustness of this chiral dynamical response. This is a key criterion for assessing the potential of non-reciprocal quantum systems for operation across diverse parametric regimes.
For fixed Rabi frequencies $\Omega_p = 6\,\gamma$ and $\Omega_c = 4\,\gamma$, we extract the oscillation frequencies ($f_{\rm cou}$, $f_{\rm co}$) and their amplitudes as a function of detuning $\Delta_c$ from the dynamical spectra for both counter- and co-propagating configurations, as summarized in Fig.~\ref{fig:4}.
The counter-propagating configuration [Fig.~\ref{fig:4}(a)] sustains synchronized dynamics over a broad detuning range from $-12.5\,\gamma$ to $4.7\,\gamma$, a behavior markedly absent in the co-propagating case (b). This demonstrates the robustness of the non-reciprocal dynamical response to parameter variations.
In non-equilibrium systems, the information can be encoded on the oscillatory frequency of transmission.
We quantify the optical non-reciprocity by the contrast ratio $\eta$, which is based on the oscillation frequencies and given by $\eta = (f_{\mathrm{cou}} - f_{\mathrm{co}}) / (f_{\mathrm{cou}} + f_{\mathrm{co}})$. Since synchronization is perfectly suppressed under co-propagation (i.e., $f_{\mathrm{co}} = 0$), a contrast ratio $\eta=100\%$ is achieved across a broad laser frequency range [see the inset of Fig.~\ref{fig:4}].

\textit{Conclusion and Discussion}.---In this work, we systematically explore non-reciprocal synchronization in a thermal Rydberg gas. Our core innovation is the introduction of motion-induced chirality, which generalizes optical non-reciprocity from the steady state to the dynamical, nonequilibrium regime~\cite{lodahl2017Chiral,zhang2018Thermalmotioninduced}, demonstrating the formation of a chiral time crystal. The robustness of this phenomenon, demonstrated by its pronounced dynamics over a wide range of laser frequency, highlights its potential for non-equilibrium quantum device design. Our findings thereby pave the way for engineering diverse non-reciprocal effects in driven-dissipative systems.


\begin{thebibliography}{34}%
	\makeatletter
	\providecommand \@ifxundefined [1]{%
		\@ifx{#1\undefined}
	}%
	\providecommand \@ifnum [1]{%
		\ifnum #1\expandafter \@firstoftwo
		\else \expandafter \@secondoftwo
		\fi
	}%
	\providecommand \@ifx [1]{%
		\ifx #1\expandafter \@firstoftwo
		\else \expandafter \@secondoftwo
		\fi
	}%
	\providecommand \natexlab [1]{#1}%
	\providecommand \enquote  [1]{``#1''}%
	\providecommand \bibnamefont  [1]{#1}%
	\providecommand \bibfnamefont [1]{#1}%
	\providecommand \citenamefont [1]{#1}%
	\providecommand \href@noop [0]{\@secondoftwo}%
	\providecommand \href [0]{\begingroup \@sanitize@url \@href}%
	\providecommand \@href[1]{\@@startlink{#1}\@@href}%
	\providecommand \@@href[1]{\endgroup#1\@@endlink}%
	\providecommand \@sanitize@url [0]{\catcode `\\12\catcode `\$12\catcode
		`\&12\catcode `\#12\catcode `\^12\catcode `\_12\catcode `\%12\relax}%
	\providecommand \@@startlink[1]{}%
	\providecommand \@@endlink[0]{}%
	\providecommand \url  [0]{\begingroup\@sanitize@url \@url }%
	\providecommand \@url [1]{\endgroup\@href {#1}{\urlprefix }}%
	\providecommand \urlprefix  [0]{URL }%
	\providecommand \Eprint [0]{\href }%
	\providecommand \doibase [0]{https://doi.org/}%
	\providecommand \selectlanguage [0]{\@gobble}%
	\providecommand \bibinfo  [0]{\@secondoftwo}%
	\providecommand \bibfield  [0]{\@secondoftwo}%
	\providecommand \translation [1]{[#1]}%
	\providecommand \BibitemOpen [0]{}%
	\providecommand \bibitemStop [0]{}%
	\providecommand \bibitemNoStop [0]{.\EOS\space}%
	\providecommand \EOS [0]{\spacefactor3000\relax}%
	\providecommand \BibitemShut  [1]{\csname bibitem#1\endcsname}%
	\let\auto@bib@innerbib\@empty
	\bibitem [{\citenamefont {Dutt}\ \emph {et~al.}(2024)\citenamefont {Dutt},
		\citenamefont {Mohanty}, \citenamefont {Gaeta},\ and\ \citenamefont
		{Lipson}}]{dutt2024Nonlinear}%
	\BibitemOpen
	\bibfield  {author} {\bibinfo {author} {\bibfnamefont {A.}~\bibnamefont
			{Dutt}}, \bibinfo {author} {\bibfnamefont {A.}~\bibnamefont {Mohanty}},
		\bibinfo {author} {\bibfnamefont {A.~L.}\ \bibnamefont {Gaeta}},\ and\
		\bibinfo {author} {\bibfnamefont {M.}~\bibnamefont {Lipson}},\ }\bibfield
	{title} {\bibinfo {title} {Nonlinear and quantum photonics using integrated
			optical materials},\ }\href {https://doi.org/10.1038/s41578-024-00668-z}
	{\bibfield  {journal} {\bibinfo  {journal} {Nat. Rev. Mater.}\ }\textbf
		{\bibinfo {volume} {9}},\ \bibinfo {pages} {321} (\bibinfo {year}
		{2024})}\BibitemShut {NoStop}%
	\bibitem [{\citenamefont {Luo}\ \emph {et~al.}(2023)\citenamefont {Luo},
		\citenamefont {Cao}, \citenamefont {Shi}, \citenamefont {Wan}, \citenamefont
		{Zhang}, \citenamefont {Li}, \citenamefont {Chen}, \citenamefont {Li},
		\citenamefont {Li}, \citenamefont {Wang}, \citenamefont {Sun}, \citenamefont
		{Karim}, \citenamefont {Cai}, \citenamefont {Kwek},\ and\ \citenamefont
		{Liu}}]{luo2023Recent}%
	\BibitemOpen
	\bibfield  {author} {\bibinfo {author} {\bibfnamefont {W.}~\bibnamefont
			{Luo}}, \bibinfo {author} {\bibfnamefont {L.}~\bibnamefont {Cao}}, \bibinfo
		{author} {\bibfnamefont {Y.}~\bibnamefont {Shi}}, \bibinfo {author}
		{\bibfnamefont {L.}~\bibnamefont {Wan}}, \bibinfo {author} {\bibfnamefont
			{H.}~\bibnamefont {Zhang}}, \bibinfo {author} {\bibfnamefont
			{S.}~\bibnamefont {Li}}, \bibinfo {author} {\bibfnamefont {G.}~\bibnamefont
			{Chen}}, \bibinfo {author} {\bibfnamefont {Y.}~\bibnamefont {Li}}, \bibinfo
		{author} {\bibfnamefont {S.}~\bibnamefont {Li}}, \bibinfo {author}
		{\bibfnamefont {Y.}~\bibnamefont {Wang}}, \bibinfo {author} {\bibfnamefont
			{S.}~\bibnamefont {Sun}}, \bibinfo {author} {\bibfnamefont {M.~F.}\
			\bibnamefont {Karim}}, \bibinfo {author} {\bibfnamefont {H.}~\bibnamefont
			{Cai}}, \bibinfo {author} {\bibfnamefont {L.~C.}\ \bibnamefont {Kwek}},\ and\
		\bibinfo {author} {\bibfnamefont {A.~Q.}\ \bibnamefont {Liu}},\ }\bibfield
	{title} {\bibinfo {title} {Recent progress in quantum photonic chips for
			quantum communication and internet},\ }\href
	{https://doi.org/10.1038/s41377-023-01173-8} {\bibfield  {journal} {\bibinfo
			{journal} {Light Sci. Appl.}\ }\textbf {\bibinfo {volume} {12}},\ \bibinfo
		{pages} {175} (\bibinfo {year} {2023})}\BibitemShut {NoStop}%
	\bibitem [{\citenamefont {Kim}\ \emph {et~al.}(2021)\citenamefont {Kim},
		\citenamefont {Sohn}, \citenamefont {Peterson},\ and\ \citenamefont
		{Bahl}}]{kim2021Onchip}%
	\BibitemOpen
	\bibfield  {author} {\bibinfo {author} {\bibfnamefont {S.}~\bibnamefont
			{Kim}}, \bibinfo {author} {\bibfnamefont {D.~B.}\ \bibnamefont {Sohn}},
		\bibinfo {author} {\bibfnamefont {C.~W.}\ \bibnamefont {Peterson}},\ and\
		\bibinfo {author} {\bibfnamefont {G.}~\bibnamefont {Bahl}},\ }\bibfield
	{title} {\bibinfo {title} {On-chip optical non-reciprocity through a
			synthetic {{Hall}} effect for photons},\ }\href
	{https://doi.org/10.1063/5.0034291} {\bibfield  {journal} {\bibinfo
			{journal} {APL Photonics}\ }\textbf {\bibinfo {volume} {6}},\ \bibinfo
		{pages} {011301} (\bibinfo {year} {2021})}\BibitemShut {NoStop}%
	\bibitem [{\citenamefont {Luo}\ \emph {et~al.}(2019)\citenamefont {Luo},
		\citenamefont {Brauner}, \citenamefont {Eigner}, \citenamefont {Sharapova},
		\citenamefont {Ricken}, \citenamefont {Meier}, \citenamefont {Herrmann},\
		and\ \citenamefont {Silberhorn}}]{luo2019Nonlinear}%
	\BibitemOpen
	\bibfield  {author} {\bibinfo {author} {\bibfnamefont {K.-H.}\ \bibnamefont
			{Luo}}, \bibinfo {author} {\bibfnamefont {S.}~\bibnamefont {Brauner}},
		\bibinfo {author} {\bibfnamefont {C.}~\bibnamefont {Eigner}}, \bibinfo
		{author} {\bibfnamefont {P.~R.}\ \bibnamefont {Sharapova}}, \bibinfo {author}
		{\bibfnamefont {R.}~\bibnamefont {Ricken}}, \bibinfo {author} {\bibfnamefont
			{T.}~\bibnamefont {Meier}}, \bibinfo {author} {\bibfnamefont
			{H.}~\bibnamefont {Herrmann}},\ and\ \bibinfo {author} {\bibfnamefont
			{C.}~\bibnamefont {Silberhorn}},\ }\bibfield  {title} {\bibinfo {title}
		{Nonlinear integrated quantum electro-optic circuits},\ }\href
	{https://doi.org/10.1126/sciadv.aat1451} {\bibfield  {journal} {\bibinfo
			{journal} {Sci. Adv.}\ }\textbf {\bibinfo {volume} {5}},\ \bibinfo {pages}
		{eaat1451} (\bibinfo {year} {2019})}\BibitemShut {NoStop}%
	\bibitem [{\citenamefont {Sounas}\ and\ \citenamefont
		{Al{\`u}}(2017)}]{sounas2017Nonreciprocal}%
	\BibitemOpen
	\bibfield  {author} {\bibinfo {author} {\bibfnamefont {D.~L.}\ \bibnamefont
			{Sounas}}\ and\ \bibinfo {author} {\bibfnamefont {A.}~\bibnamefont
			{Al{\`u}}},\ }\bibfield  {title} {\bibinfo {title} {Non-reciprocal photonics
			based on time modulation},\ }\href
	{https://doi.org/10.1038/s41566-017-0051-x} {\bibfield  {journal} {\bibinfo
			{journal} {Nat. Photon.}\ }\textbf {\bibinfo {volume} {11}},\ \bibinfo
		{pages} {774} (\bibinfo {year} {2017})}\BibitemShut {NoStop}%
	\bibitem [{\citenamefont {Shen}\ \emph {et~al.}(2016)\citenamefont {Shen},
		\citenamefont {Zhang}, \citenamefont {Chen}, \citenamefont {Zou},
		\citenamefont {Xiao}, \citenamefont {Zou}, \citenamefont {Sun}, \citenamefont
		{Guo},\ and\ \citenamefont {Dong}}]{shen2016Experimental}%
	\BibitemOpen
	\bibfield  {author} {\bibinfo {author} {\bibfnamefont {Z.}~\bibnamefont
			{Shen}}, \bibinfo {author} {\bibfnamefont {Y.-L.}\ \bibnamefont {Zhang}},
		\bibinfo {author} {\bibfnamefont {Y.}~\bibnamefont {Chen}}, \bibinfo {author}
		{\bibfnamefont {C.-L.}\ \bibnamefont {Zou}}, \bibinfo {author} {\bibfnamefont
			{Y.-F.}\ \bibnamefont {Xiao}}, \bibinfo {author} {\bibfnamefont {X.-B.}\
			\bibnamefont {Zou}}, \bibinfo {author} {\bibfnamefont {F.-W.}\ \bibnamefont
			{Sun}}, \bibinfo {author} {\bibfnamefont {G.-C.}\ \bibnamefont {Guo}},\ and\
		\bibinfo {author} {\bibfnamefont {C.-H.}\ \bibnamefont {Dong}},\ }\bibfield
	{title} {\bibinfo {title} {Experimental realization of optomechanically
			induced non-reciprocity},\ }\href {https://doi.org/10.1038/nphoton.2016.161}
	{\bibfield  {journal} {\bibinfo  {journal} {Nat. Photon.}\ }\textbf {\bibinfo
			{volume} {10}},\ \bibinfo {pages} {657} (\bibinfo {year} {2016})}\BibitemShut
	{NoStop}%
	\bibitem [{\citenamefont {Khanikaev}\ and\ \citenamefont
		{Al{\`u}}(2015)}]{khanikaev2015Nonlinear}%
	\BibitemOpen
	\bibfield  {author} {\bibinfo {author} {\bibfnamefont {A.~B.}\ \bibnamefont
			{Khanikaev}}\ and\ \bibinfo {author} {\bibfnamefont {A.}~\bibnamefont
			{Al{\`u}}},\ }\bibfield  {title} {\bibinfo {title} {Nonlinear dynamic
			reciprocity},\ }\href {https://doi.org/10.1038/nphoton.2015.86} {\bibfield
		{journal} {\bibinfo  {journal} {Nat. Photon.}\ }\textbf {\bibinfo {volume}
			{9}},\ \bibinfo {pages} {359} (\bibinfo {year} {2015})}\BibitemShut {NoStop}%
	\bibitem [{\citenamefont {Lodahl}\ \emph {et~al.}(2017)\citenamefont {Lodahl},
		\citenamefont {Mahmoodian}, \citenamefont {Stobbe}, \citenamefont
		{Rauschenbeutel}, \citenamefont {Schneeweiss}, \citenamefont {Volz},
		\citenamefont {Pichler},\ and\ \citenamefont {Zoller}}]{lodahl2017Chiral}%
	\BibitemOpen
	\bibfield  {author} {\bibinfo {author} {\bibfnamefont {P.}~\bibnamefont
			{Lodahl}}, \bibinfo {author} {\bibfnamefont {S.}~\bibnamefont {Mahmoodian}},
		\bibinfo {author} {\bibfnamefont {S.}~\bibnamefont {Stobbe}}, \bibinfo
		{author} {\bibfnamefont {A.}~\bibnamefont {Rauschenbeutel}}, \bibinfo
		{author} {\bibfnamefont {P.}~\bibnamefont {Schneeweiss}}, \bibinfo {author}
		{\bibfnamefont {J.}~\bibnamefont {Volz}}, \bibinfo {author} {\bibfnamefont
			{H.}~\bibnamefont {Pichler}},\ and\ \bibinfo {author} {\bibfnamefont
			{P.}~\bibnamefont {Zoller}},\ }\bibfield  {title} {\bibinfo {title} {Chiral
			quantum optics},\ }\href {https://doi.org/10.1038/nature21037} {\bibfield
		{journal} {\bibinfo  {journal} {Nature}\ }\textbf {\bibinfo {volume} {541}},\
		\bibinfo {pages} {473} (\bibinfo {year} {2017})}\BibitemShut {NoStop}%
	\bibitem [{\citenamefont {Huang}\ \emph {et~al.}(2020)\citenamefont {Huang},
		\citenamefont {Cenker}, \citenamefont {Zhang}, \citenamefont {Ray},
		\citenamefont {Song}, \citenamefont {Taniguchi}, \citenamefont {Watanabe},
		\citenamefont {McGuire}, \citenamefont {Xiao},\ and\ \citenamefont
		{Xu}}]{huang2020Tuning}%
	\BibitemOpen
	\bibfield  {author} {\bibinfo {author} {\bibfnamefont {B.}~\bibnamefont
			{Huang}}, \bibinfo {author} {\bibfnamefont {J.}~\bibnamefont {Cenker}},
		\bibinfo {author} {\bibfnamefont {X.}~\bibnamefont {Zhang}}, \bibinfo
		{author} {\bibfnamefont {E.~L.}\ \bibnamefont {Ray}}, \bibinfo {author}
		{\bibfnamefont {T.}~\bibnamefont {Song}}, \bibinfo {author} {\bibfnamefont
			{T.}~\bibnamefont {Taniguchi}}, \bibinfo {author} {\bibfnamefont
			{K.}~\bibnamefont {Watanabe}}, \bibinfo {author} {\bibfnamefont {M.~A.}\
			\bibnamefont {McGuire}}, \bibinfo {author} {\bibfnamefont {D.}~\bibnamefont
			{Xiao}},\ and\ \bibinfo {author} {\bibfnamefont {X.}~\bibnamefont {Xu}},\
	}\bibfield  {title} {\bibinfo {title} {Tuning inelastic light scattering via
			symmetry control in the two-dimensional magnet {{CrI3}}},\ }\href
	{https://doi.org/10.1038/s41565-019-0598-4} {\bibfield  {journal} {\bibinfo
			{journal} {Nat. Nanotechnol.}\ }\textbf {\bibinfo {volume} {15}},\ \bibinfo
		{pages} {212} (\bibinfo {year} {2020})}\BibitemShut {NoStop}%
	\bibitem [{\citenamefont {Feng}\ \emph {et~al.}(2020)\citenamefont {Feng},
		\citenamefont {Hanke}, \citenamefont {Zhou}, \citenamefont {Guo},
		\citenamefont {Bl{\"u}gel}, \citenamefont {Mokrousov},\ and\ \citenamefont
		{Yao}}]{feng2020Topological}%
	\BibitemOpen
	\bibfield  {author} {\bibinfo {author} {\bibfnamefont {W.}~\bibnamefont
			{Feng}}, \bibinfo {author} {\bibfnamefont {J.-P.}\ \bibnamefont {Hanke}},
		\bibinfo {author} {\bibfnamefont {X.}~\bibnamefont {Zhou}}, \bibinfo {author}
		{\bibfnamefont {G.-Y.}\ \bibnamefont {Guo}}, \bibinfo {author} {\bibfnamefont
			{S.}~\bibnamefont {Bl{\"u}gel}}, \bibinfo {author} {\bibfnamefont
			{Y.}~\bibnamefont {Mokrousov}},\ and\ \bibinfo {author} {\bibfnamefont
			{Y.}~\bibnamefont {Yao}},\ }\bibfield  {title} {\bibinfo {title} {Topological
			magneto-optical effects and their quantization in noncoplanar
			antiferromagnets},\ }\href {https://doi.org/10.1038/s41467-019-13968-8}
	{\bibfield  {journal} {\bibinfo  {journal} {Nat. Commun.}\ }\textbf {\bibinfo
			{volume} {11}},\ \bibinfo {pages} {118} (\bibinfo {year} {2020})}\BibitemShut
	{NoStop}%
	\bibitem [{\citenamefont {Wang}\ \emph {et~al.}(2025)\citenamefont {Wang},
		\citenamefont {Zhang}, \citenamefont {Hu}, \citenamefont {Chen},
		\citenamefont {Li}, \citenamefont {Yang}, \citenamefont {Zou}, \citenamefont
		{Zhang}, \citenamefont {Dong}, \citenamefont {Li}, \citenamefont {Zhang},
		\citenamefont {Guo},\ and\ \citenamefont {Zou}}]{wang2025SelfinducedOptical}%
	\BibitemOpen
	\bibfield  {author} {\bibinfo {author} {\bibfnamefont {Z.-B.}\ \bibnamefont
			{Wang}}, \bibinfo {author} {\bibfnamefont {Y.-L.}\ \bibnamefont {Zhang}},
		\bibinfo {author} {\bibfnamefont {X.-X.}\ \bibnamefont {Hu}}, \bibinfo
		{author} {\bibfnamefont {G.-J.}\ \bibnamefont {Chen}}, \bibinfo {author}
		{\bibfnamefont {M.}~\bibnamefont {Li}}, \bibinfo {author} {\bibfnamefont
			{P.-F.}\ \bibnamefont {Yang}}, \bibinfo {author} {\bibfnamefont {X.-B.}\
			\bibnamefont {Zou}}, \bibinfo {author} {\bibfnamefont {P.-F.}\ \bibnamefont
			{Zhang}}, \bibinfo {author} {\bibfnamefont {C.-H.}\ \bibnamefont {Dong}},
		\bibinfo {author} {\bibfnamefont {G.}~\bibnamefont {Li}}, \bibinfo {author}
		{\bibfnamefont {T.-C.}\ \bibnamefont {Zhang}}, \bibinfo {author}
		{\bibfnamefont {G.-C.}\ \bibnamefont {Guo}},\ and\ \bibinfo {author}
		{\bibfnamefont {C.-L.}\ \bibnamefont {Zou}},\ }\bibfield  {title} {\bibinfo
		{title} {Self-induced optical non-reciprocity},\ }\href
	{https://doi.org/10.1038/s41377-024-01692-y} {\bibfield  {journal} {\bibinfo
			{journal} {Light Sci. Appl.}\ }\textbf {\bibinfo {volume} {14}},\ \bibinfo
		{pages} {23} (\bibinfo {year} {2025})}\BibitemShut {NoStop}%
	\bibitem [{\citenamefont {Liang}\ \emph {et~al.}(2020)\citenamefont {Liang},
		\citenamefont {Liu}, \citenamefont {Xu}, \citenamefont {Wen}, \citenamefont
		{Lu}, \citenamefont {Xia}, \citenamefont {Tey}, \citenamefont {Liu},\ and\
		\citenamefont {You}}]{liang2020CollisionInduced}%
	\BibitemOpen
	\bibfield  {author} {\bibinfo {author} {\bibfnamefont {C.}~\bibnamefont
			{Liang}}, \bibinfo {author} {\bibfnamefont {B.}~\bibnamefont {Liu}}, \bibinfo
		{author} {\bibfnamefont {A.-N.}\ \bibnamefont {Xu}}, \bibinfo {author}
		{\bibfnamefont {X.}~\bibnamefont {Wen}}, \bibinfo {author} {\bibfnamefont
			{C.}~\bibnamefont {Lu}}, \bibinfo {author} {\bibfnamefont {K.}~\bibnamefont
			{Xia}}, \bibinfo {author} {\bibfnamefont {M.~K.}\ \bibnamefont {Tey}},
		\bibinfo {author} {\bibfnamefont {Y.-C.}\ \bibnamefont {Liu}},\ and\ \bibinfo
		{author} {\bibfnamefont {L.}~\bibnamefont {You}},\ }\bibfield  {title}
	{\bibinfo {title} {Collision-{{Induced Broadband Optical Nonreciprocity}}},\
	}\href {https://doi.org/10.1103/PhysRevLett.125.123901} {\bibfield  {journal}
		{\bibinfo  {journal} {Phys. Rev. Lett.}\ }\textbf {\bibinfo {volume} {125}},\
		\bibinfo {pages} {123901} (\bibinfo {year} {2020})}\BibitemShut {NoStop}%
	\bibitem [{\citenamefont {Zhang}\ \emph {et~al.}(2018)\citenamefont {Zhang},
		\citenamefont {Hu}, \citenamefont {Lin}, \citenamefont {Niu}, \citenamefont
		{Xia}, \citenamefont {Gong},\ and\ \citenamefont
		{Gong}}]{zhang2018Thermalmotioninduced}%
	\BibitemOpen
	\bibfield  {author} {\bibinfo {author} {\bibfnamefont {S.}~\bibnamefont
			{Zhang}}, \bibinfo {author} {\bibfnamefont {Y.}~\bibnamefont {Hu}}, \bibinfo
		{author} {\bibfnamefont {G.}~\bibnamefont {Lin}}, \bibinfo {author}
		{\bibfnamefont {Y.}~\bibnamefont {Niu}}, \bibinfo {author} {\bibfnamefont
			{K.}~\bibnamefont {Xia}}, \bibinfo {author} {\bibfnamefont {J.}~\bibnamefont
			{Gong}},\ and\ \bibinfo {author} {\bibfnamefont {S.}~\bibnamefont {Gong}},\
	}\bibfield  {title} {\bibinfo {title} {Thermal-motion-induced non-reciprocal
			quantum optical system},\ }\href {https://doi.org/10.1038/s41566-018-0269-2}
	{\bibfield  {journal} {\bibinfo  {journal} {Nat. Photon.}\ }\textbf {\bibinfo
			{volume} {12}},\ \bibinfo {pages} {744} (\bibinfo {year} {2018})}\BibitemShut
	{NoStop}%
	\bibitem [{\citenamefont {Huang}\ \emph {et~al.}(2018)\citenamefont {Huang},
		\citenamefont {Miranowicz}, \citenamefont {Liao}, \citenamefont {Nori},\ and\
		\citenamefont {Jing}}]{huang2018Nonreciprocal}%
	\BibitemOpen
	\bibfield  {author} {\bibinfo {author} {\bibfnamefont {R.}~\bibnamefont
			{Huang}}, \bibinfo {author} {\bibfnamefont {A.}~\bibnamefont {Miranowicz}},
		\bibinfo {author} {\bibfnamefont {J.-Q.}\ \bibnamefont {Liao}}, \bibinfo
		{author} {\bibfnamefont {F.}~\bibnamefont {Nori}},\ and\ \bibinfo {author}
		{\bibfnamefont {H.}~\bibnamefont {Jing}},\ }\bibfield  {title} {\bibinfo
		{title} {Nonreciprocal {{Photon Blockade}}},\ }\href
	{https://doi.org/10.1103/PhysRevLett.121.153601} {\bibfield  {journal}
		{\bibinfo  {journal} {Phys. Rev. Lett.}\ }\textbf {\bibinfo {volume} {121}},\
		\bibinfo {pages} {153601} (\bibinfo {year} {2018})}\BibitemShut {NoStop}%
	\bibitem [{\citenamefont {Zola}\ \emph {et~al.}(2019)\citenamefont {Zola},
		\citenamefont {Bisoyi}, \citenamefont {Wang}, \citenamefont {Urbas},
		\citenamefont {Bunning},\ and\ \citenamefont {Li}}]{zola2019Dynamic}%
	\BibitemOpen
	\bibfield  {author} {\bibinfo {author} {\bibfnamefont {R.~S.}\ \bibnamefont
			{Zola}}, \bibinfo {author} {\bibfnamefont {H.~K.}\ \bibnamefont {Bisoyi}},
		\bibinfo {author} {\bibfnamefont {H.}~\bibnamefont {Wang}}, \bibinfo {author}
		{\bibfnamefont {A.~M.}\ \bibnamefont {Urbas}}, \bibinfo {author}
		{\bibfnamefont {T.~J.}\ \bibnamefont {Bunning}},\ and\ \bibinfo {author}
		{\bibfnamefont {Q.}~\bibnamefont {Li}},\ }\bibfield  {title} {\bibinfo
		{title} {Dynamic {{Control}} of {{Light Direction Enabled}} by
			{{Stimuli}}-{{Responsive Liquid Crystal Gratings}}},\ }\href
	{https://doi.org/10.1002/adma.201806172} {\bibfield  {journal} {\bibinfo
			{journal} {Adv. Mater.}\ }\textbf {\bibinfo {volume} {31}},\ \bibinfo {pages}
		{1806172} (\bibinfo {year} {2019})}\BibitemShut {NoStop}%
	\bibitem [{\citenamefont {Dorrah}\ \emph {et~al.}(2021)\citenamefont {Dorrah},
		\citenamefont {Rubin}, \citenamefont {Zaidi}, \citenamefont {Tamagnone},\
		and\ \citenamefont {Capasso}}]{dorrah2021Metasurface}%
	\BibitemOpen
	\bibfield  {author} {\bibinfo {author} {\bibfnamefont {A.~H.}\ \bibnamefont
			{Dorrah}}, \bibinfo {author} {\bibfnamefont {N.~A.}\ \bibnamefont {Rubin}},
		\bibinfo {author} {\bibfnamefont {A.}~\bibnamefont {Zaidi}}, \bibinfo
		{author} {\bibfnamefont {M.}~\bibnamefont {Tamagnone}},\ and\ \bibinfo
		{author} {\bibfnamefont {F.}~\bibnamefont {Capasso}},\ }\bibfield  {title}
	{\bibinfo {title} {Metasurface optics for on-demand polarization
			transformations along the optical path},\ }\href
	{https://doi.org/10.1038/s41566-020-00750-2} {\bibfield  {journal} {\bibinfo
			{journal} {Nat. Photon.}\ }\textbf {\bibinfo {volume} {15}},\ \bibinfo
		{pages} {287} (\bibinfo {year} {2021})}\BibitemShut {NoStop}%
	\bibitem [{\citenamefont {Jin}(2018)}]{jin2018IncidentDirection}%
	\BibitemOpen
	\bibfield  {author} {\bibinfo {author} {\bibfnamefont {L.}~\bibnamefont
			{Jin}},\ }\bibfield  {title} {\bibinfo {title} {Incident {{Direction
					Independent Wave Propagation}} and {{Unidirectional Lasing}}},\ }\href
	{https://doi.org/10.1103/PhysRevLett.121.073901} {\bibfield  {journal}
		{\bibinfo  {journal} {Phys. Rev. Lett.}\ }\textbf {\bibinfo {volume} {121}},\
		\bibinfo {pages} {073901} (\bibinfo {year} {2018})}\BibitemShut {NoStop}%
	\bibitem [{\citenamefont {Xie}\ \emph {et~al.}(2025)\citenamefont {Xie},
		\citenamefont {Sun}, \citenamefont {Wu}, \citenamefont {Li}, \citenamefont
		{Guo}, \citenamefont {Yi},\ and\ \citenamefont {Xiang}}]{xie2025Chiral}%
	\BibitemOpen
	\bibfield  {author} {\bibinfo {author} {\bibfnamefont {C.}~\bibnamefont
			{Xie}}, \bibinfo {author} {\bibfnamefont {K.}~\bibnamefont {Sun}}, \bibinfo
		{author} {\bibfnamefont {K.-D.}\ \bibnamefont {Wu}}, \bibinfo {author}
		{\bibfnamefont {C.-F.}\ \bibnamefont {Li}}, \bibinfo {author} {\bibfnamefont
			{G.-C.}\ \bibnamefont {Guo}}, \bibinfo {author} {\bibfnamefont
			{W.}~\bibnamefont {Yi}},\ and\ \bibinfo {author} {\bibfnamefont {G.-Y.}\
			\bibnamefont {Xiang}},\ }\bibfield  {title} {\bibinfo {title} {Chiral
			switching of many-body steady states in a dissipative {{Rydberg}} gas},\
	}\href {https://doi.org/10.1016/j.scib.2025.08.051} {\bibfield  {journal}
		{\bibinfo  {journal} {Sci. Bull.}\ ,\ \bibinfo {pages} {in press}} (\bibinfo
		{year} {2025})}\BibitemShut {NoStop}%
	\bibitem [{\citenamefont
		{Wadenpfuhl}(2023)}]{wadenpfuhl2023EmergenceSynchronization}%
	\BibitemOpen
	\bibfield  {author} {\bibinfo {author} {\bibfnamefont {K.}~\bibnamefont
			{Wadenpfuhl}},\ }\bibfield  {title} {\bibinfo {title} {Emergence of
			{{Synchronization}} in a {{Driven-Dissipative Hot Rydberg Vapor}}},\ }\href
	{https://doi.org/10.1103/PhysRevLett.131.143002} {\bibfield  {journal}
		{\bibinfo  {journal} {Phys. Rev. Lett.}\ }\textbf {\bibinfo {volume} {131}},\
		\bibinfo {pages} {143002} (\bibinfo {year} {2023})}\BibitemShut {NoStop}%
	\bibitem [{\citenamefont {Wu}\ \emph {et~al.}(2024)\citenamefont {Wu},
		\citenamefont {Wang}, \citenamefont {Yang}, \citenamefont {Gao},
		\citenamefont {Liang}, \citenamefont {Tey}, \citenamefont {Li}, \citenamefont
		{Pohl},\ and\ \citenamefont {You}}]{wu2024Dissipative}%
	\BibitemOpen
	\bibfield  {author} {\bibinfo {author} {\bibfnamefont {X.}~\bibnamefont
			{Wu}}, \bibinfo {author} {\bibfnamefont {Z.}~\bibnamefont {Wang}}, \bibinfo
		{author} {\bibfnamefont {F.}~\bibnamefont {Yang}}, \bibinfo {author}
		{\bibfnamefont {R.}~\bibnamefont {Gao}}, \bibinfo {author} {\bibfnamefont
			{C.}~\bibnamefont {Liang}}, \bibinfo {author} {\bibfnamefont {M.~K.}\
			\bibnamefont {Tey}}, \bibinfo {author} {\bibfnamefont {X.}~\bibnamefont
			{Li}}, \bibinfo {author} {\bibfnamefont {T.}~\bibnamefont {Pohl}},\ and\
		\bibinfo {author} {\bibfnamefont {L.}~\bibnamefont {You}},\ }\bibfield
	{title} {\bibinfo {title} {Dissipative time crystal in a strongly interacting
			{{Rydberg}} gas},\ }\href {https://doi.org/10.1038/s41567-024-02542-9}
	{\bibfield  {journal} {\bibinfo  {journal} {Nat. Phys.}\ }\textbf {\bibinfo
			{volume} {20}},\ \bibinfo {pages} {1389} (\bibinfo {year}
		{2024})}\BibitemShut {NoStop}%
	\bibitem [{\citenamefont {Ding}\ \emph {et~al.}(2024)\citenamefont {Ding},
		\citenamefont {Bai}, \citenamefont {Liu}, \citenamefont {Shi}, \citenamefont
		{Guo}, \citenamefont {Li},\ and\ \citenamefont {Adams}}]{ding2024Ergodicity}%
	\BibitemOpen
	\bibfield  {author} {\bibinfo {author} {\bibfnamefont {D.}~\bibnamefont
			{Ding}}, \bibinfo {author} {\bibfnamefont {Z.}~\bibnamefont {Bai}}, \bibinfo
		{author} {\bibfnamefont {Z.}~\bibnamefont {Liu}}, \bibinfo {author}
		{\bibfnamefont {B.}~\bibnamefont {Shi}}, \bibinfo {author} {\bibfnamefont
			{G.}~\bibnamefont {Guo}}, \bibinfo {author} {\bibfnamefont {W.}~\bibnamefont
			{Li}},\ and\ \bibinfo {author} {\bibfnamefont {C.~S.}\ \bibnamefont
			{Adams}},\ }\bibfield  {title} {\bibinfo {title} {Ergodicity breaking from
			{{Rydberg}} clusters in a driven-dissipative many-body system},\ }\href
	{https://doi.org/10.1126/sciadv.adl5893} {\bibfield  {journal} {\bibinfo
			{journal} {Sci. Adv.}\ }\textbf {\bibinfo {volume} {10}},\ \bibinfo {pages}
		{eadl5893} (\bibinfo {year} {2024})}\BibitemShut {NoStop}%
	\bibitem [{\citenamefont {Bernien}\ \emph {et~al.}(2017)\citenamefont
		{Bernien}, \citenamefont {Schwartz}, \citenamefont {Keesling}, \citenamefont
		{Levine}, \citenamefont {Omran}, \citenamefont {Pichler}, \citenamefont
		{Choi}, \citenamefont {Zibrov}, \citenamefont {Endres}, \citenamefont
		{Greiner}, \citenamefont {Vuleti{\'c}},\ and\ \citenamefont
		{Lukin}}]{bernien2017Probing}%
	\BibitemOpen
	\bibfield  {author} {\bibinfo {author} {\bibfnamefont {H.}~\bibnamefont
			{Bernien}}, \bibinfo {author} {\bibfnamefont {S.}~\bibnamefont {Schwartz}},
		\bibinfo {author} {\bibfnamefont {A.}~\bibnamefont {Keesling}}, \bibinfo
		{author} {\bibfnamefont {H.}~\bibnamefont {Levine}}, \bibinfo {author}
		{\bibfnamefont {A.}~\bibnamefont {Omran}}, \bibinfo {author} {\bibfnamefont
			{H.}~\bibnamefont {Pichler}}, \bibinfo {author} {\bibfnamefont
			{S.}~\bibnamefont {Choi}}, \bibinfo {author} {\bibfnamefont {A.~S.}\
			\bibnamefont {Zibrov}}, \bibinfo {author} {\bibfnamefont {M.}~\bibnamefont
			{Endres}}, \bibinfo {author} {\bibfnamefont {M.}~\bibnamefont {Greiner}},
		\bibinfo {author} {\bibfnamefont {V.}~\bibnamefont {Vuleti{\'c}}},\ and\
		\bibinfo {author} {\bibfnamefont {M.~D.}\ \bibnamefont {Lukin}},\ }\bibfield
	{title} {\bibinfo {title} {Probing many-body dynamics on a 51-atom quantum
			simulator},\ }\href {https://doi.org/10.1038/nature24622} {\bibfield
		{journal} {\bibinfo  {journal} {Nature}\ }\textbf {\bibinfo {volume} {551}},\
		\bibinfo {pages} {579} (\bibinfo {year} {2017})}\BibitemShut {NoStop}%
	\bibitem [{\citenamefont {Adams}\ \emph {et~al.}(2020)\citenamefont {Adams},
		\citenamefont {Pritchard},\ and\ \citenamefont {Shaffer}}]{adams2020Rydberg}%
	\BibitemOpen
	\bibfield  {author} {\bibinfo {author} {\bibfnamefont {C.~S.}\ \bibnamefont
			{Adams}}, \bibinfo {author} {\bibfnamefont {J.~D.}\ \bibnamefont
			{Pritchard}},\ and\ \bibinfo {author} {\bibfnamefont {J.~P.}\ \bibnamefont
			{Shaffer}},\ }\bibfield  {title} {\bibinfo {title} {Rydberg atom quantum
			technologies},\ }\href {https://doi.org/10.1088/1361-6455/ab52ef} {\bibfield
		{journal} {\bibinfo  {journal} {J. Phys. B}\ }\textbf {\bibinfo {volume}
			{53}},\ \bibinfo {pages} {012002} (\bibinfo {year} {2020})}\BibitemShut
	{NoStop}%
	\bibitem [{\citenamefont {Bendkowsky}\ \emph {et~al.}(2009)\citenamefont
		{Bendkowsky}, \citenamefont {Butscher}, \citenamefont {Nipper}, \citenamefont
		{Shaffer}, \citenamefont {L{\"o}w},\ and\ \citenamefont
		{Pfau}}]{bendkowsky2009Observation}%
	\BibitemOpen
	\bibfield  {author} {\bibinfo {author} {\bibfnamefont {V.}~\bibnamefont
			{Bendkowsky}}, \bibinfo {author} {\bibfnamefont {B.}~\bibnamefont
			{Butscher}}, \bibinfo {author} {\bibfnamefont {J.}~\bibnamefont {Nipper}},
		\bibinfo {author} {\bibfnamefont {J.~P.}\ \bibnamefont {Shaffer}}, \bibinfo
		{author} {\bibfnamefont {R.}~\bibnamefont {L{\"o}w}},\ and\ \bibinfo {author}
		{\bibfnamefont {T.}~\bibnamefont {Pfau}},\ }\bibfield  {title} {\bibinfo
		{title} {Observation of ultralong-range {{Rydberg}} molecules},\ }\href
	{https://doi.org/10.1038/nature07945} {\bibfield  {journal} {\bibinfo
			{journal} {Nature}\ }\textbf {\bibinfo {volume} {458}},\ \bibinfo {pages}
		{1005} (\bibinfo {year} {2009})}\BibitemShut {NoStop}%
	\bibitem [{\citenamefont {Saffman}\ \emph {et~al.}(2010)\citenamefont
		{Saffman}, \citenamefont {Walker},\ and\ \citenamefont
		{M{\o}lmer}}]{saffman2010Quantum}%
	\BibitemOpen
	\bibfield  {author} {\bibinfo {author} {\bibfnamefont {M.}~\bibnamefont
			{Saffman}}, \bibinfo {author} {\bibfnamefont {T.~G.}\ \bibnamefont
			{Walker}},\ and\ \bibinfo {author} {\bibfnamefont {K.}~\bibnamefont
			{M{\o}lmer}},\ }\bibfield  {title} {\bibinfo {title} {Quantum information
			with {{Rydberg}} atoms},\ }\href {https://doi.org/10.1103/RevModPhys.82.2313}
	{\bibfield  {journal} {\bibinfo  {journal} {Rev. Mod. Phys.}\ }\textbf
		{\bibinfo {volume} {82}},\ \bibinfo {pages} {2313} (\bibinfo {year}
		{2010})}\BibitemShut {NoStop}%
	\bibitem [{\citenamefont {Ebadi}\ \emph {et~al.}(2021)\citenamefont {Ebadi},
		\citenamefont {Wang}, \citenamefont {Levine}, \citenamefont {Keesling},
		\citenamefont {Semeghini}, \citenamefont {Omran}, \citenamefont {Bluvstein},
		\citenamefont {Samajdar}, \citenamefont {Pichler}, \citenamefont {Ho},
		\citenamefont {Choi}, \citenamefont {Sachdev}, \citenamefont {Greiner},
		\citenamefont {Vuleti{\'c}},\ and\ \citenamefont
		{Lukin}}]{ebadi2021Quantuma}%
	\BibitemOpen
	\bibfield  {author} {\bibinfo {author} {\bibfnamefont {S.}~\bibnamefont
			{Ebadi}}, \bibinfo {author} {\bibfnamefont {T.~T.}\ \bibnamefont {Wang}},
		\bibinfo {author} {\bibfnamefont {H.}~\bibnamefont {Levine}}, \bibinfo
		{author} {\bibfnamefont {A.}~\bibnamefont {Keesling}}, \bibinfo {author}
		{\bibfnamefont {G.}~\bibnamefont {Semeghini}}, \bibinfo {author}
		{\bibfnamefont {A.}~\bibnamefont {Omran}}, \bibinfo {author} {\bibfnamefont
			{D.}~\bibnamefont {Bluvstein}}, \bibinfo {author} {\bibfnamefont
			{R.}~\bibnamefont {Samajdar}}, \bibinfo {author} {\bibfnamefont
			{H.}~\bibnamefont {Pichler}}, \bibinfo {author} {\bibfnamefont {W.~W.}\
			\bibnamefont {Ho}}, \bibinfo {author} {\bibfnamefont {S.}~\bibnamefont
			{Choi}}, \bibinfo {author} {\bibfnamefont {S.}~\bibnamefont {Sachdev}},
		\bibinfo {author} {\bibfnamefont {M.}~\bibnamefont {Greiner}}, \bibinfo
		{author} {\bibfnamefont {V.}~\bibnamefont {Vuleti{\'c}}},\ and\ \bibinfo
		{author} {\bibfnamefont {M.~D.}\ \bibnamefont {Lukin}},\ }\bibfield  {title}
	{\bibinfo {title} {Quantum phases of matter on a 256-atom programmable
			quantum simulator},\ }\href {https://doi.org/10.1038/s41586-021-03582-4}
	{\bibfield  {journal} {\bibinfo  {journal} {Nature}\ }\textbf {\bibinfo
			{volume} {595}},\ \bibinfo {pages} {227} (\bibinfo {year}
		{2021})}\BibitemShut {NoStop}%
	\bibitem [{\citenamefont {{Guardado-Sanchez}}\ \emph
		{et~al.}(2018)\citenamefont {{Guardado-Sanchez}}, \citenamefont {Brown},
		\citenamefont {Mitra}, \citenamefont {Devakul}, \citenamefont {Huse},
		\citenamefont {Schau{\ss}},\ and\ \citenamefont
		{Bakr}}]{guardado-sanchez2018Probing}%
	\BibitemOpen
	\bibfield  {author} {\bibinfo {author} {\bibfnamefont {E.}~\bibnamefont
			{{Guardado-Sanchez}}}, \bibinfo {author} {\bibfnamefont {P.~T.}\ \bibnamefont
			{Brown}}, \bibinfo {author} {\bibfnamefont {D.}~\bibnamefont {Mitra}},
		\bibinfo {author} {\bibfnamefont {T.}~\bibnamefont {Devakul}}, \bibinfo
		{author} {\bibfnamefont {D.~A.}\ \bibnamefont {Huse}}, \bibinfo {author}
		{\bibfnamefont {P.}~\bibnamefont {Schau{\ss}}},\ and\ \bibinfo {author}
		{\bibfnamefont {W.~S.}\ \bibnamefont {Bakr}},\ }\bibfield  {title} {\bibinfo
		{title} {Probing the {{Quench Dynamics}} of {{Antiferromagnetic
					Correlations}} in a {{2D Quantum Ising Spin System}}},\ }\href
	{https://doi.org/10.1103/PhysRevX.8.021069} {\bibfield  {journal} {\bibinfo
			{journal} {Phys. Rev. X}\ }\textbf {\bibinfo {volume} {8}},\ \bibinfo {pages}
		{021069} (\bibinfo {year} {2018})}\BibitemShut {NoStop}%
	\bibitem [{\citenamefont {Turner}\ \emph {et~al.}(2018)\citenamefont {Turner},
		\citenamefont {Michailidis}, \citenamefont {Abanin}, \citenamefont {Serbyn},\
		and\ \citenamefont {Papi{\'c}}}]{turner2018Weak}%
	\BibitemOpen
	\bibfield  {author} {\bibinfo {author} {\bibfnamefont {C.~J.}\ \bibnamefont
			{Turner}}, \bibinfo {author} {\bibfnamefont {A.~A.}\ \bibnamefont
			{Michailidis}}, \bibinfo {author} {\bibfnamefont {D.~A.}\ \bibnamefont
			{Abanin}}, \bibinfo {author} {\bibfnamefont {M.}~\bibnamefont {Serbyn}},\
		and\ \bibinfo {author} {\bibfnamefont {Z.}~\bibnamefont {Papi{\'c}}},\
	}\bibfield  {title} {\bibinfo {title} {Weak ergodicity breaking from quantum
			many-body scars},\ }\href {https://doi.org/10.1038/s41567-018-0137-5}
	{\bibfield  {journal} {\bibinfo  {journal} {Nat. Phys.}\ }\textbf {\bibinfo
			{volume} {14}},\ \bibinfo {pages} {745} (\bibinfo {year} {2018})}\BibitemShut
	{NoStop}%
	\bibitem [{\citenamefont {Bluvstein}\ \emph {et~al.}(2021)\citenamefont
		{Bluvstein}, \citenamefont {Omran}, \citenamefont {Levine}, \citenamefont
		{Keesling}, \citenamefont {Semeghini}, \citenamefont {Ebadi}, \citenamefont
		{Wang}, \citenamefont {Michailidis}, \citenamefont {Maskara}, \citenamefont
		{Ho}, \citenamefont {Choi}, \citenamefont {Serbyn}, \citenamefont {Greiner},
		\citenamefont {Vuleti{\'c}},\ and\ \citenamefont
		{Lukin}}]{bluvstein2021Controlling}%
	\BibitemOpen
	\bibfield  {author} {\bibinfo {author} {\bibfnamefont {D.}~\bibnamefont
			{Bluvstein}}, \bibinfo {author} {\bibfnamefont {A.}~\bibnamefont {Omran}},
		\bibinfo {author} {\bibfnamefont {H.}~\bibnamefont {Levine}}, \bibinfo
		{author} {\bibfnamefont {A.}~\bibnamefont {Keesling}}, \bibinfo {author}
		{\bibfnamefont {G.}~\bibnamefont {Semeghini}}, \bibinfo {author}
		{\bibfnamefont {S.}~\bibnamefont {Ebadi}}, \bibinfo {author} {\bibfnamefont
			{T.~T.}\ \bibnamefont {Wang}}, \bibinfo {author} {\bibfnamefont {A.~A.}\
			\bibnamefont {Michailidis}}, \bibinfo {author} {\bibfnamefont
			{N.}~\bibnamefont {Maskara}}, \bibinfo {author} {\bibfnamefont {W.~W.}\
			\bibnamefont {Ho}}, \bibinfo {author} {\bibfnamefont {S.}~\bibnamefont
			{Choi}}, \bibinfo {author} {\bibfnamefont {M.}~\bibnamefont {Serbyn}},
		\bibinfo {author} {\bibfnamefont {M.}~\bibnamefont {Greiner}}, \bibinfo
		{author} {\bibfnamefont {V.}~\bibnamefont {Vuleti{\'c}}},\ and\ \bibinfo
		{author} {\bibfnamefont {M.~D.}\ \bibnamefont {Lukin}},\ }\bibfield  {title}
	{\bibinfo {title} {Controlling quantum many-body dynamics in driven
			{{Rydberg}} atom arrays},\ }\href {https://doi.org/10.1126/science.abg2530}
	{\bibfield  {journal} {\bibinfo  {journal} {Science}\ }\textbf {\bibinfo
			{volume} {371}},\ \bibinfo {pages} {1355} (\bibinfo {year}
		{2021})}\BibitemShut {NoStop}%
	\bibitem [{\citenamefont {Lee}\ \emph {et~al.}(2011)\citenamefont {Lee},
		\citenamefont {H{\"a}ffner},\ and\ \citenamefont
		{Cross}}]{lee2011AntiferromagneticPhase}%
	\BibitemOpen
	\bibfield  {author} {\bibinfo {author} {\bibfnamefont {T.~E.}\ \bibnamefont
			{Lee}}, \bibinfo {author} {\bibfnamefont {H.}~\bibnamefont {H{\"a}ffner}},\
		and\ \bibinfo {author} {\bibfnamefont {M.~C.}\ \bibnamefont {Cross}},\
	}\bibfield  {title} {\bibinfo {title} {Antiferromagnetic phase transition in
			a nonequilibrium lattice of {{Rydberg}} atoms},\ }\href
	{https://doi.org/10.1103/physreva.84.031402} {\bibfield  {journal} {\bibinfo
			{journal} {Phys. Rev. A}\ }\textbf {\bibinfo {volume} {84}},\ \bibinfo
		{pages} {031402} (\bibinfo {year} {2011})}\BibitemShut {NoStop}%
	\bibitem [{\citenamefont {Carr}\ \emph {et~al.}(2013)\citenamefont {Carr},
		\citenamefont {Ritter}, \citenamefont {Wade}, \citenamefont {Adams},\ and\
		\citenamefont {Weatherill}}]{carr2013Nonequilibrium}%
	\BibitemOpen
	\bibfield  {author} {\bibinfo {author} {\bibfnamefont {C.}~\bibnamefont
			{Carr}}, \bibinfo {author} {\bibfnamefont {R.}~\bibnamefont {Ritter}},
		\bibinfo {author} {\bibfnamefont {C.~G.}\ \bibnamefont {Wade}}, \bibinfo
		{author} {\bibfnamefont {C.~S.}\ \bibnamefont {Adams}},\ and\ \bibinfo
		{author} {\bibfnamefont {K.~J.}\ \bibnamefont {Weatherill}},\ }\bibfield
	{title} {\bibinfo {title} {Nonequilibrium {{Phase Transition}} in a {{Dilute
					Rydberg Ensemble}}},\ }\href {https://doi.org/10.1103/PhysRevLett.111.113901}
	{\bibfield  {journal} {\bibinfo  {journal} {Phys. Rev. Lett.}\ }\textbf
		{\bibinfo {volume} {111}},\ \bibinfo {pages} {113901} (\bibinfo {year}
		{2013})}\BibitemShut {NoStop}%
	\bibitem [{\citenamefont {Bai}\ \emph {et~al.}(2020)\citenamefont {Bai},
		\citenamefont {Adams}, \citenamefont {Huang},\ and\ \citenamefont
		{Li}}]{bai2020SelfInduced}%
	\BibitemOpen
	\bibfield  {author} {\bibinfo {author} {\bibfnamefont {Z.}~\bibnamefont
			{Bai}}, \bibinfo {author} {\bibfnamefont {C.~S.}\ \bibnamefont {Adams}},
		\bibinfo {author} {\bibfnamefont {G.}~\bibnamefont {Huang}},\ and\ \bibinfo
		{author} {\bibfnamefont {W.}~\bibnamefont {Li}},\ }\bibfield  {title}
	{\bibinfo {title} {Self-{{Induced Transparency}} in {{Warm}} and {{Strongly
					Interacting Rydberg Gases}}},\ }\href
	{https://doi.org/10.1103/PhysRevLett.125.263605} {\bibfield  {journal}
		{\bibinfo  {journal} {Phys. Rev. Lett.}\ }\textbf {\bibinfo {volume} {125}},\
		\bibinfo {pages} {263605} (\bibinfo {year} {2020})}\BibitemShut {NoStop}%
	\bibitem [{\citenamefont {Kuramoto}(1975)}]{kuramoto1975international}%
	\BibitemOpen
	\bibfield  {author} {\bibinfo {author} {\bibfnamefont {Y.}~\bibnamefont
			{Kuramoto}},\ }\bibfield  {title} {\bibinfo {title} {International symposium
			on mathematical problems in theoretical physics},\ }\href@noop {} {\bibfield
		{journal} {\bibinfo  {journal} {Lect. Notes Phys.}\ }\textbf {\bibinfo
			{volume} {30}},\ \bibinfo {pages} {420} (\bibinfo {year} {1975})}\BibitemShut
	{NoStop}%
	\bibitem [{\citenamefont {Acebr{\'o}n}\ \emph {et~al.}(2005)\citenamefont
		{Acebr{\'o}n}, \citenamefont {Bonilla}, \citenamefont {P{\'e}rez~Vicente},
		\citenamefont {Ritort},\ and\ \citenamefont {Spigler}}]{acebron2005Kuramoto}%
	\BibitemOpen
	\bibfield  {author} {\bibinfo {author} {\bibfnamefont {J.~A.}\ \bibnamefont
			{Acebr{\'o}n}}, \bibinfo {author} {\bibfnamefont {L.~L.}\ \bibnamefont
			{Bonilla}}, \bibinfo {author} {\bibfnamefont {C.~J.}\ \bibnamefont
			{P{\'e}rez~Vicente}}, \bibinfo {author} {\bibfnamefont {F.}~\bibnamefont
			{Ritort}},\ and\ \bibinfo {author} {\bibfnamefont {R.}~\bibnamefont
			{Spigler}},\ }\bibfield  {title} {\bibinfo {title} {The {{Kuramoto}} model:
			{{A}} simple paradigm for synchronization phenomena},\ }\href
	{https://doi.org/10.1103/RevModPhys.77.137} {\bibfield  {journal} {\bibinfo
			{journal} {Rev. Mod. Phys.}\ }\textbf {\bibinfo {volume} {77}},\ \bibinfo
		{pages} {137} (\bibinfo {year} {2005})}\BibitemShut {NoStop}%
\end{thebibliography}

%

\end{document}